# The Vocabulary of Introductory Physics and Its Implications for Learning Physics

Salomon F. Itza-Ortiz, N. Sanjay Rebello and Dean Zollman
*Kansas State University, Department of Physics, Physics Education Research Group, Manhattan KS, 66506*
and
Manuel Rodriguez-Achach
*Universidad Autonoma de Yucatan, Fac. de Ingenieria, Ave. Industrias nocontaminantes, A.P. 150 Cordemex, Yucatan, Mexico, C.P. 97310*

Abstract:

We investigate students' use of words in everyday language as well as in physics We find students are more likely to identify and explain the meaning of the word as it is used in physics when they have become familiar with the parameters involved in the physics concept.

## I. Introduction

In science new words might be "invented" to name or describe new processes, discoveries or inventions. However, for the most part, the scientific vocabulary is formed from words we use throughout our lives in everyday language. When we begin studying science we learn new meanings of words we had previously used. Sometimes these new meanings may contradict everyday meanings or seem counterintuitive. We often learn words in association with objects and situations.[1] Due to these associations that students bring to class, they may not interpret the physics meaning correctly. This misinterpretation of language leads students to confusion that is sometimes classified as a misconception.[2-6] Research about the semantics used in physics textbooks[7-9] and the meaning of words has been done,[10-12] but the problem seems to go beyond semantics.[8]

The linguistic relativity hypothesis, sometimes referred to as the Sapir-Whorf hypothesis,[1] says that "we see and hear and otherwise experience very largely as we do because the language habits of our community predispose certain choices of interpretation." An upshot of this hypothesis is that *language may not determine thought, but it certainly may influence thought.*[1] We have to make students conscious of the fact that though the words may remain the same, their everyday meaning is no longer a figure of speech, but a technical meaning (physics meaning). That is, we need to change the way students may "think" about words. In spite of the close relationship between language and thought, most research does not address the semantics used in physics textbooks[7-9] and the meaning of words[10-12]. This study, however, will address that relationship.

In this paper we present results of research done at Kansas State University and at the Universidad Autonoma de Yucatan in Mexico. We provide insights on the implications of the use of everyday language in the learning of physics concepts. Our main question is: Do the differences in the use of words between everyday life and physics inhibit learning of physics? We focus on three words that are common in any introductory physics course: "force," "momentum" and "impulse." The following sections describe our research goals, methodology and results. In our conclusion we provide some suggestions to help students incorporate the physics meaning of these words into their vocabulary.

II. Research Goals and Methodology

Our goal was to study how students perceive the similarities and differences between the everyday and physics meanings of the words "force," "momentum" and "impulse." We were also interested in studying whether these perceived differences and similarities affect the learning of those concepts in physics. A major portion of the data was collected at Kansas State. Our research participants consisted of 154 students enrolled in "The Physical World I" course, a course that is taken by non-science majors, most of who are in their junior year. "Conceptual Physics" by Paul Hewitt is the text for the course. Fifty-seven percent of the students had previously taken at least one physics course.

Through a collaborator at the Universidad Autonoma de Yucatan in Mexico we carried out a component of our research with native Spanish speakers. We would like to know if Spanish-speaking students have similar problems to English-speakers, in using their vocabulary for the word "force." Because of schedule conflicts we only research this word.

Our research at Kansas State consisted of three phases: (1) pre-survey, (2) post-survey and (3) interview. All 154 students participated in the surveys and 14 were selected for interviews. In the pre-survey we asked students to make up three different sentences using the word "force" or variants from it. The term "force" had not been introduced in class at the time of the pre-survey. Thus, it showed how the students would use the word in their everyday vocabulary. We sorted the sentences into four classifications according to the usage of the word "force:" Verb Animate, when the word is used as a verb and relates to a subject (person or animal); Verb Inanimate, when the word is used as a verb and relates to an inanimate object; Noun, when the word is used as a noun; and Adjective or Adverb, when the corresponding variant of the word is used as an adjective or adverb. Table I shows the most frequently written sentences of each type.

The post-survey was administered after the term "force" was introduced in class. For this survey, we chose four sentences from the list and presented them to the students. We asked students to explain the similarities and differences between the use of the word "force" in the given sentence and its use in physics. The results from the second survey were classified into three categories: Category 1 included students who can explain how the word "force," as used in each of the sentences, is both similar to and different from the word "force" as used in physics; Category 2 included students who are able to describe these similarities and differences for only a few of the given sentences; and Category 3 included students whose responses indicate they cannot explain these similarities and differences for any of the given sentences. The categorization of students' sentences was validated by an independent researcher. Immediately after the post survey the course instructor administered a scheduled class test that evaluated course material and included the concept of force. In our analysis we focused on the score for the questions relevant only to force --- 9 out of a total of 26 multiple choice questions. Only two of these questions required numerical calculations, the other seven questions were conceptual. These conceptual questions were similar to the ones in the Force Concept Inventory (FCI).[13] This post-survey was translated to Spanish and applied to freshman engineering students at Universidad Autonoma de Yucatan in Mexico.

The interview protocol at Kansas State was similar to the procedure followed on the written surveys. Each student was first asked to write a sentence using the word "force," then she or he was asked to explain how the meaning of the word "force" as used in their sentence is similar and dissimilar to the word as used in physics. Later they were presented with a few selected sentences containing the word "force" and asked the same questions.

We followed an identical survey protocol (all three stages), for the words "momentum" and "impulse." For the results from the second survey of these words, we had only two classifications, Noun and Adjective. Table II shows the most frequently written sentences of each. Immediately after the corresponding second survey the course instructor administered the scheduled class test, which evaluated these two concepts among others. Because of the course structure the number of questions on these concepts was reduced to six by the course instructor. Thus, to have significance we combined the results from these two words. The questions were multiple-choice, three requiring simple numerical calculations and three of the conceptual type. The interviews on these words followed the same protocol as the interview about the word "force."

### III. Results and Discussion

#### A. The word force

Fifty-nine percent of sentences on the pre-survey included the word "force" as a verb. This observation is consistent with the fact that the word "force" is often used as a verb in everyday language.[11,12] Thirty-six percent of the students in the second survey were in categories 1 and 2, *i.e.* they described the similarities and differences between the meaning of the word "force" in the given sentences and its physics meaning. The remaining 64% of the students, category 3, are apparently not able to differentiate between the everyday and physics meaning of the word "force." Fig. 1 shows the "cumulative frequency" curves for categories 1, 2 and 3 on the second survey vs. the students' test scores. From it, 80% (80th percentile) of the students in category 1 have grades below 91, the same percentages of students in category 2 have grades below 89, and from category 3, grades below 84. Thus, students who could identify and explain the physics meaning of the word "force" obtained better test scores. We believe this establish a link between the linguistic ability of students to discern various meanings of the word "force" and their conceptual understanding of the concept of force, as measured by the test. To further probe our results we interviewed 14 students, individually, using representatives from each of the three categories. The students first wrote two sentences using the word "force" (or its derivative). They were asked to think aloud about their sentences and describe whether the ways in which they had used the word "force" were similar or different from the ways in which they used it in everyday life. All students were able to identify if the way they were using the word "force" had an everyday or physics meaning. When asked why the word "force" in one of their sentences would have a physics meaning, they responded by stating that the word relates to pushing, pulling or motion. When asked why it would have an everyday meaning, they said it has to do with mental power, power, or following rules --- not in a physical sense. Their explanation for the physics meaning is consistent with what they were taught in class: force is "any influence that tends to accelerate an object; a push or a pull." They also were taught that force equals mass times acceleration. Only two out of the 14 students were able to relate force to the mass of the object and/or its acceleration. In the second part of the interview the students were given four sentences and asked to identify the meaning of the word "force" in each sentence. All students were able to identify whether the meaning corresponded to everyday life or physics because they focused on the context of the sentence. However, almost all students were unable to explain how the meaning of the word is similar or different to its physics meaning. Only two students who identified force with mass and/or acceleration were able to explain the similarities and differences of the meaning of force in the sentence with its meaning in physics. Thus, all students were able to explain whether the word "force" in the sentences has an everyday or physics meaning, but only those who identified the parameters associated with force were able to explain how the word "force" in the sentence was similar and different to its use in physics. For example, when a student was asked to explain the meaning of the word force in physics, he/she said "*Force is weight, force of a book onto a table; force of a person while pushing a chair across the room.*" When this same student was asked to explain the meaning of the word "force" in the sentence "The bulldozer forced the rock into the ditch," he/she said, "*the bulldozer has direct contact onto* [*sic*] *the rock, pushes the rock.*" He/she identified force as a push, from the definition of force. Another student stated that, "*Force causes movement, there are forces everywhere, like friction. Force is mass times acceleration.*" When this student was asked to explain the meaning of the word "force" in the bulldozer sentence, he/she said "*the bulldozer moves the rock into place, there is mass and acceleration.*" This last student is using the parameters involved in force to explain why the word force in the sentence has a physics meaning. He/she is attempting to assimilate the meaning of the word.

We obtained some interesting results from the surveys given to undergraduates in Mexico. We found that Spanish-speaking students in Mexico are very likely to use the word "fuerza" (force) as a synonym for "poder" (power) in the sense of "ability to act or produce an effect," *i.e.* they use the word "force" as a verb ("forzar"), which is similar to the way English-speaking students at Kansas State responded to the same survey. We believe this similarity in results is because the word "fuerza" is spelled similar to the word "forzar," which is a verb --- in English the word "force" is used both as noun and verb. Thus, it seems that students are more familiar with the verb usage, the everyday meaning. In terms of the Sapir-Whorf hypothesis,[1] the language habits predispose a choice of interpretation. We infer from this that it is possible that if we had administered these surveys to Italian-speaking students, where the words are "forza," for noun, and "forzare" for verb, we would have found the same results. In contrast, we would expect that students with a native language where the noun and verb forms of the word "force" are different, they could make the distinction. For example, in German "kraft" is force as a noun, and "erzwingen" is force as a verb.

B.  The words momentum and impulse

Momentum and impulse were discussed in class after the topic of force. Eighty percent of the sentences written by the students used the words "momentum" and "impulse" as a noun in the pre-survey. This is consistent with the common usage of these words in everyday language. On the post-survey 36% of the students were placed in categories 1 and 2. That is, they were able to differentiate between the everyday and the physics meaning of the words and explain the physics meaning. This is the same percentage of students that resulted in the post-survey for the word "force," although our records indicate they are not the same group of students. Fig. 2 shows the "cumulative frequency" curves for categories 1, 2 and 3 on the second survey vs. the students' test scores on questions pertaining to the concepts of momentum or impulse. Eighty percent ($80^{th}$ percentile) of the students in category 1 have grades below 90, the same percentages of students in category 2 have grades below 80, and from category 3 grades below 60. In general students in category 1 score higher on the test than students in categories 2 and 3. These results are very similar to the ones for the word "force," reinforcing the idea of a link between the linguistic ability of students to discern various meanings of a word and their conceptual understanding of the concept described by the word. The difficulty with these words also showed in the test scores of the students.

The participants in the interview phase were the same students as before. We asked them to write two sentences using the word "momentum" and two with "impulse." Twelve students interpreted the meaning of the word "momentum" in the physics context. However, only six of them related momentum to mass and/or velocity. When asked to explain momentum in physics, typical answers included terms such as "the mass of the object, speed, action, motion, or build up of energy." When relating to an everyday meaning, the students said momentum had to do with feelings or mental action, not physical motion. It is interesting to note that momentum has a Latin root which means "movement," so this word by itself relates to motion. The everyday meaning of the term is quite similar to its physics meaning. It appears that due to this similarity of meanings, students are more likely to explain the physics meaning of the term momentum. For instance, when asked to explain the meaning of this term, one student said, "*When someone is running, he has mass and speed, he is creating momentum.*" Another said, "*Momentum in physics is, ... as something falls speed up. In a slope gains speed, gains momentum.*" For the word "impulse" only one of the students was able to explain the meaning of the term impulse as used in physics. The other students used the everyday meaning of the term. They said impulse has to do with instant action, spontaneity, or something you do without thinking about it. The dictionary meaning of the word "impulse" is a sudden spontaneous inclination or incitement to some usually unpremeditated action. This word is very well embedded in students' minds and it is difficult for them to relate it to physics. In fact the physics meaning of the term, product of the force acting and the time duration for which it acts, is quite different from the everyday meaning. It appears that this difference makes it difficult for students to understand its physics meaning. From the two students quoted above, when they were asked to explain the meaning of the term impulse, the first student said, "*Impulse is something involuntarily, it just happens.*" The second student said, "*impulse is a force, a push, ... not sure.*" The first student is describing the everyday meaning, but the second student, albeit doubtfully, is relating impulse to force. This is the only student who did not relate impulse to instantaneous actions. Thus, the word "momentum" seems more intuitive to the students. They might not define it as velocity times mass but they always relate it to motion. The word "impulse" is not as intuitive to the students, because its everyday meaning is quite different from its meaning in physics. Again the Sapir-Whorf hypothesis[1] seems to be applicable here since it is the everyday meaning of these words that is the most influential in students' thoughts.

IV. Impact on Instruction

One way the acquisition of knowledge can be conceptualized is through the idea that students acquire different understandings of relevant concepts. These coexist and compete with previous informal understandings.[14] We propose the idea that comparing everyday and physics meanings of words will help students to assimilate the meaning of the word in physics. When making these comparisons the students can relate to the parameters involved in the physics term, thus helping them to establish connections between the words and building their "physics vocabulary." We do not believe the physics meaning of words will take the place of the everyday meaning, but rather they would coexist. Students can be asked to compare the physics and everyday meanings of the words by writing essays in different contexts.[15] Many of the students in conceptual physics classes, such as humanities majors, have strong writing abilities and may find such tasks to be quite enjoyable. Efforts to inculcate superior writing skills across the curriculum have been used in several high schools and colleges. The writing exercises described above may be helpful in such a curriculum.

V. Conclusions

We surveyed a physics class with non-science majors to study students' perceptions of the similarities between the everyday and physics meanings of three commonly used words. Our findings show that students who can differentiate between the everyday and physics meanings of the words and, can explain the physics meaning are more likely to obtain higher test scores. From interviews we conjecture that students who are able to identify or remember parameters related to the word are more likely to explain its physics meaning. We translated and administered one of the surveys on the word "force" to Spanish-speaking undergraduates in Mexico. Our results indicate that even in other languages where the word "force" can be used as verb or noun, students are more likely to use it in an everyday connotation. For the other two words included in our study, "momentum" and "impulse," we found that the word "momentum" seems more intuitive to the students, as they always related it to motion. The word "impulse" is not as intuitive to the students; its everyday meaning is quite different from its meaning in physics. Unfortunately the idea of looking at differences between everyday and physics meanings of the words seems not to be carried out from force to momentum and impulse, the tests scores for the last two words are lower (Figs. 1 and 2). Our results also indicate that physics instructors should be more cognizant of the use of language and the alternative meanings of physics terminology that their students bring with them to class. We also propose that instructors can devise special writing assignments that would enable students to overcome this linguistic barrier in learning physics.

## Acknowledgments

This work was supported, in part, by the National Science Foundation under grant # REC-0087788. We would also like to thank Professor Emmett Wright for his valuable comments and Dr. Seunghee Lee for the validation of surveys and results.

Figures and Tables

Figure 1

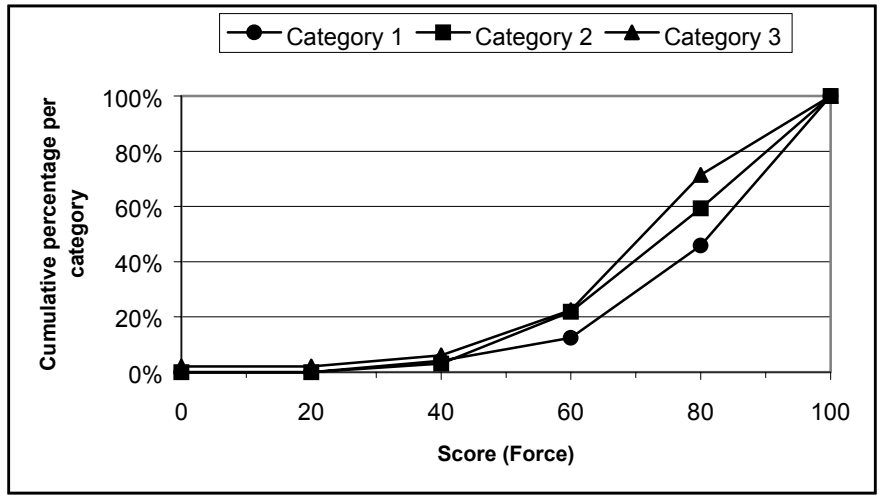

Figure 2

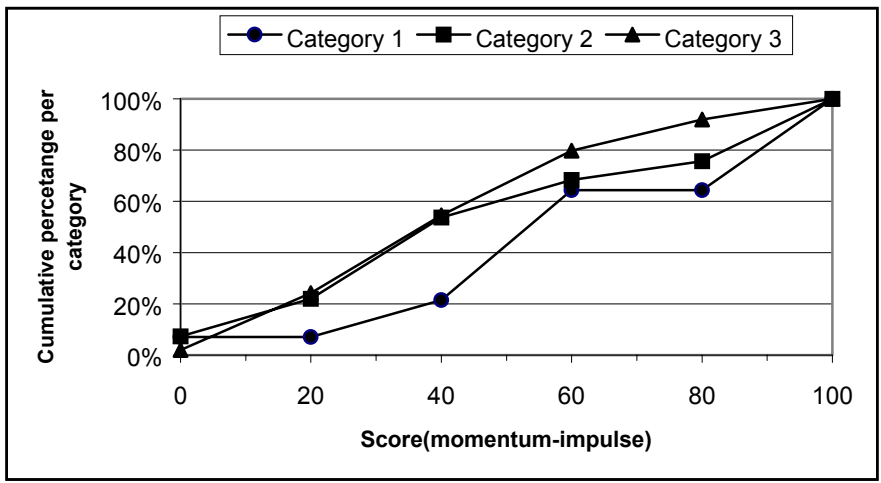

Table I

| Classification | Sentence |
|---|---|
| **Verb Inanimate** | "I forced the box into the closet." |
| | "Jim was forcing the nut on the bolt." |
| **Verb Animate** | "I forced myself to go to class everyday." |
| | "My parents forced me to go to college." |
| **Noun** | "The force on the ball made it move." |
| | "The bomb exploded with great force." |
| | "I was hit by the force of the 18 wheeler." |
| **Adjective** | "She used a very forceful tone of voice." |

Table II

| Classification | Sentence |
|---|---|
| **Noun** | "After their touchdown, the other team had the momentum." |
| | "The football player had a lot of momentum when he tackled his opponent." |
| | "Our team gained momentum in the game after intercepting the ball." |
| | "As the car rolled down the hill it gained momentum." |
| | "An impulse made her change her mind" |
| | "My first impulse was to kick him." |
| | "In time of crisis we act on our impulses." |
| **Adjective** | "My sister is an impulsive shopper." |

Captions

Figure 1: Cumulative frequency curve for students' test scores on the word "force." At the 80$^{th}$ percentile students in category 1 have scores below 91, category 2 score 89 and category 3 score 84. Thus, students who can identify and explain the physics meaning of the word "force" (category 1) obtain higher test scores.

Figure 2: Cumulative frequency curve for students' test scores on the words "momentum" and "impulse." At the 80$^{th}$ percentile students in category 1 have scores below 90, category 2 score 80 and category 3 score 60. Thus, students who can identify and explain the physics meaning of the words "momentum" and "impulse" (category 1) obtain higher test scores.

Table I: Classification of sentences collected from students with the word "force" or a derivative from it. The students are more likely to use force as a verb.

Table II: Classification of sentences collected from students with the words "momentum" and "impulse." The students use these words only as nouns or adjectives.